\documentclass[superscriptaddress, aps, pra, amssymb, amsfonts, 
twocolumn]{revtex4}
\pdfoutput=1
\usepackage{amsmath, amssymb, bbm, changes}
\usepackage{epstopdf}
\usepackage[all]{xy}
\usepackage{ifpdf}
\ifpdf
 \usepackage[pdftex]{graphicx}
 \else
\usepackage{graphicx}
 \fi
\usepackage{graphicx}
\newcommand{\unity}{\ensuremath{\mathbbm 1}}

\def\<{\langle}
\def\>{\rangle}

\def\hcal{{\cal H}}

\newcommand{\be}{\begin{eqnarray} \begin{aligned}}
\newcommand{\ee}{\end{aligned} \end{eqnarray} }
\newcommand{\benn}{\begin{eqnarray*} \begin{aligned}}
\newcommand{\eenn}{\end{aligned} \end{eqnarray*} }

\newcommand{\ben}{\begin{eqnarray} \begin{aligned}}
\newcommand{\een}{\end{aligned} \end{eqnarray} }

\newcommand{\bc}{\begin{center}}
\newcommand{\ec}{\end{center}}

\newcommand{\e}{\mathrm{e}}

\newcommand{\beq}{\begin{eqnarray} \begin{aligned}}
\newcommand{\eeq}{\end{aligned} \end{eqnarray} }
\newcommand{\bea}{\begin{array}}
\newcommand{\eea}{\end{array}}

\newcommand{\bee}{\begin{enumerate}}
\newcommand{\eee}{\end{enumerate}}
\newcommand{\bei}{\begin{itemize}}
\newcommand{\eei}{\end{itemize}}

\usepackage{amsfonts}

\def\01{\{0,1\}}

\newcommand{\jono}[1]{\textcolor{blue}{#1}}

\newcommand{\supl}{Supplementary Information}

\def\<{\langle}
\def\>{\rangle}

\def\hcal{{\cal H}}

\def\s{\,\,\,\,}

\def\trumpd\mathsf{  D_{work}}

\def\E{\mathrm{e}}

\def\s{_\mathsf{S}}

\def\b{_\mathsf{B}}
\def\w{w_{\mathsf{max}}}
\newtheorem{result}{Result}
\begin{document}
\title{A derivation (and quantification) of the third law of thermodynamics}
\author{Llu\'{\i}s \surname{Masanes}}
\author{Jonathan \surname{Oppenheim}}
\affiliation{University College of London, Department of Physics \& Astronomy, London, WC1E 6BT and London Interdisciplinary Network for Quantum Science}   
\begin{abstract}
The third law of thermodynamics has a controversial past and a number of formulations due to Planck, Einstein, and Nernst. It's most
accepted version, the unattainability principle~\cite{Nernst-1912}, states that {\it any thermodynamic process cannot reach the temperature of absolute zero by a finite number of steps and within a finite time.}
Although formulated in 1912, there has been no general proof of the principle, and the only evidence we have for it is that particular cooling methods become less efficient as the temperature decreases.
Here we provide the first derivation of a general unattainability principle, which applies to arbitrary cooling processes, even those exploiting the laws of quantum mechanics or involving an infinite-dimensional reservoir.
We quantify the resources needed to cool a system to any particular temperature, and translate these resources into a minimal time or number of steps by considering the notion of a Thermal Machine which obeys similar restrictions to universal computers. We generally find that the obtainable temperature can scale as an inverse power of the cooling time.
Our argument relies on the heat capacity of the bath being positive, and we show that if this is not the case then perfect cooling in finite time is in principle possible. 
Our results also clarify the connection between two versions of the third law (the Unattainability Principle and the Heat Theorem), 
and place ultimate bounds on the speed at which information can be erased.
\end{abstract}
\maketitle

\textit{Introduction.-} Walther Nernst's first formulation of the third law of thermodynamics~\cite{Nernst-1906}, now called the Heat Theorem, was the subject of intense discussion~\cite{kox2006confusion}. Nernst claimed that he could prove his Heat Thoerem using thermodynamical arguments 
while Einstein, who refuted several versions of Nernst's attempted derivation, 
was convinced that classical thermodynamics was not sufficient for a proof, and that quantum theory had to be taken into account. 
Max Planck's formulation~\cite{planck-third}: {\it when the temperature of a pure substance approaches absolute zero, its entropy approaches zero};
may hold for many crystalline substances, but it is not true in general, and formulations due to Einstein~\cite{einstein-third} 
and Nernst~\cite{Nernst2-1906}
were for some time considered to be typically true from an experimental point of view, but sometimes violated. 

\begin{figure}[h]
\includegraphics[width=0.39\textwidth]
{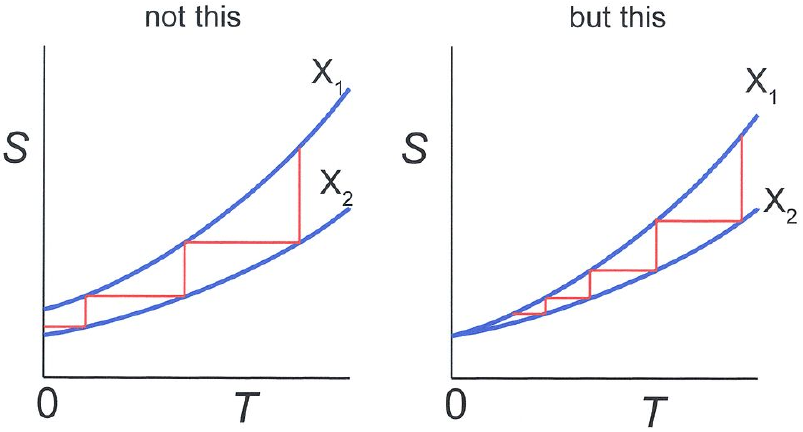}
\caption{Nernst argued that if the Heat Theorem  is violated then perfect cooling can be achieved with a finite number steps.
On the right, absolute zero is reached after an infinite number of isothermic and adiabatic reversible processes, when the Heat Theorem is satisfied $S(0,x_1) = S(0, x_2)$. 
While on the left, a finite number of steps appears to be suficient when the Heat Theorem is satisfied $S(0,x_1)= S(0,x_2)$. 
The problem with this argument is that the last adiabatic is impossible, because it must preserve the probability distribution set by the last isotherm, which is not confined to the ground-space. 
[Figure, courtesy Wikipedia Foundation]}
\label{fig:wiki}
\end{figure}

A modern understanding of entropy and quantum theory takes the Heat Theorem outside the realm of thermodynamics.
Nernst's version 
states that: {\it at zero temperature, a finite size
system has an entropy $S$ which is independent of any external paramaters $x$}, that is $S(T,x_1)-S(T,x_2)\rightarrow 0$ as the temperature $T\rightarrow 0$.
Since we now understand the entropy at zero temperature to be the logarithm of the ground state degeneracy, the validity of the Heat Theorem is contingent on 
whether the degeneracy changes for different parameters of the Hamiltonian. One can easily find families of Hamiltonians which satisfy or violate the Heat Theorem
\cite{leff1970proof, aizenman1981third}. 
Here however, we concern ourselves with the question of Nernst's Unattainability Principle~\cite{Nernst-1912}, that Nernst introduced
to support his attempted derivations of his Heat Theorem and counter Einstein's objections. We can understand it as saying that
putting a system into its ground state requires infinite time or an infinite number of steps. 
Nernst argues that if the Heat Theorem could be violated, then it would be possible to violate the Unattainability Principle (see Figure~\ref{fig:wiki}). We will see that this is not the case. 
Although one can potentially cool at a faster rate in systems violating the Heat Theorem, we show that the Unattainability Principle still holds. 
The bound we obtain is able to quantify the extent to which a change in entropy at $T=0$ affects the cooling rate.

Independently of this debate, the validity of the Unattainability Principle has remained open. This despite the central importance that cooling has in enabling quanutm phenomena in optical, atomic, and condensed matter systems.  
Quantum computation, precision measurements, quantum simulations, and the manipulation of materials at the atomic scale, all rely on extreme cooling.
Currently, we only know that certain cooling protocols, whether it be laser cooling, algorithmic cooling, dynamic cooling or the traditional alternating adaibatic and isothermic reversible operations, require
infinite time to cool a system to absolute zero.
The analysis of particular cooling protocols~\cite{levy2012quantum,allahverdyan2013comment} yields quantitative bounds on how fast cooling can take place, provided one makes certain physical assumptions. 
While for other protocols and physical assumptions,
there are claims of violations of the third law~\cite{PhysRevLett.108.120603}, followed by 
counter-claims~\cite{levy2012comments} and counter-counter-claims~\cite{kolavr2012quantum}. 
The limitation of these results is that, certain physical assumptions may not be valid at arbitrarily low temperature, or that certain protocols may not be optimal. Without a proof based on first principles, the validity of the 
third law is in question and cannot be held in the same esteem as the other laws of thermodynamics. 
A number of recent works analyze a process closely related to cooling to absolute zero: erasing information or generating pure states.
In~\cite{ticozzi2014quantum, wu2013no, HMJ} it is shown that, regardless of the amount of time invested, these processes are strictly impossible if the reservoir is finite-dimensional.  
However, strict unattainability in the sense of Nernst is not really a physically meaningful statement. Rather, one wants to obtain a finite bound to how close one can get to the desired state with a finite amount of 
resources or within a given time.
Some interesting steps in this direction are taken in~\cite{reeb2013proving}, where they obtain a bound in terms of the dimension of the reservoir, but not one that can be translated into time. 
Hence it also requires the dimension of the reservoir to be finite, an assumption that is not needed to derive our unattainability result here, and something which rarely holds in real setups. 
In fact, we shall see that the physical mechanism which enforces our third law is not dimension, but the profile of the density of states of the reservoir.
\cite{browne2014publisher} on the other hand, argues that for a qubit, one can produce a pure state to arbitrary accuracy as the time invested increases. 
This, however, requires that the work injected in the bath fluctuates around its mean value by an unbounded amount (this is also necessary in~\cite{reeb2013proving}). A fact that becomes more relevant when cooling systems much larger than a qubit. 

In the present article we bound the achievable temperature by resources such as the volume of the reservoir and the
largest value by which the work can fluctuate. This in itself can be said to constitute a version of the third law. However, we also argue that, in any process implemented in finite time, 
these two resources must remain finite too.
When the scaling of these resources with time has a standard form (explained below), and the heat bath consists of radiation, our third law provides the following relation between
the lowest achievable temperature $T'\s$ and time $t$
\begin{equation}
  T'\s \geq \frac {\mathsf{const}}{t^7}\ .
  \label{eq:inversepoly}
\end{equation}
We believe this to be the first derivation of a quantitative lower bound on the temperature in terms of cooling resources, 
confirming the general validity of the models and conjecture in~\cite{levy2012quantum} (although since we do not require the system to be continually
thermal, we are able to get a bound which is more general than the differential equation postulated there). 


The question of how much time a state transformation takes is a very natural question to ask in the field of theoretical computer science, which generally tries to quantify the resources needed to perform a task. In the
case of lower bounding the resources to perform a computation, this can be in terms of the number of basic steps or gates (given a certain energy).
It is thus no surprise, that the techniques we will use come from recent efforts to construct a theory of thermodynamics based on fundamental principles of 
quantum information theory~\cite{uniqueinfo, 
Beth-thermo, 
workvalue, popescu2006entanglement, gemmer2009quantum, del2011thermodynamic, popescu2006entanglement, 
HO-limitations, 
aaberg-singleshot, 
thermoiid, faist2012quantitative, Skrzypczyk13, brandao2013second, cwiklinski2014limitations, lostaglio2014thermodynamic, halpern2014beyond, wilming2014weak, mueller2014work}.
Traditionally, thermodynamics has been mostly concerned with large, classical systems, but these recent results also apply to 
microscopic quantum systems in arbitrary non-equilibrium states coupled to a heat reservoir, as is the case here. We thus wish to
contribute to the program of deriving the whole of thermodynamics from more fundamental principles.

\medskip
{\it Quantifying cooling resources.-} Our goal is to provide ultimate quantitative bounds applicable to any cooling procedure -- namely, we wish to find a lower bound for the temperature that a system can reach after any process 
which uses some given resources or lasting some given time $t$.
Therefore, we must allow for the most general quantum transformation i.e. those that (i) respect total energy conservation and (ii) are microscopically reversible (unitary).
This general setup includes thermodynamically irreversible protocols and also 
unrealistic protocols where total control of the microscopic degrees of freedom of the bath is required.
Surprisingly, we will find here, as was found for the case of the second law~\cite{thermoiid,HO-limitations, brandao2013second,Skrzypczyk13}, that having such unrealistic degree of control does not appear to give one an advantage over having very crude control. 

We will show that the density of states of the reservoir assisting the cooling process has an important impact on how fast a system can be cooled.
(The density of states $\Omega (E)$ is the number of states with energy $E$.)
We see that the faster $\Omega (E)$ grows, 
the lower the temperature that can be achieved with fixed resources or in a fixed amount of time. Even more: if $\Omega (E)$ grows exponentially or faster, then cooling to absolute zero in finite time is in principle possible, allowing for a violation of the third law.
However, exponential or super-exponential $\Omega (E)$ is regarded as unphysical, since it does not allow for having a well-defined thermal state and partition function at all temperatures. In more physical terms, such a substance would suck up all the energy from the surrounding systems (for certain initial conditions), and also, it would violate Beckenstein's entropy bound~\cite{bekenstein1981universal}. 
A possible concern of the reader may be that $\Omega (E)$ is often approximated by an exponential, but this is because the reservoir is often assumed to have infinite volume $V$.  Here we argue that in any process implemented in finite time, the system can only interact with a finite region of a bath.
All the above becomes more intuitive when expressed in terms of the (micro-canonical) heat capacity $C(E)$, related to $S(E) = \ln\Omega (E)$ via
\begin{equation}
  \label{mic HC}
  C(E) = 
  -\frac {[S'(E)]^2} {S''(E)}
  \ ,
\end{equation}
where primes represent differentials.
If $S(E)$ grows linearly or faster, then $C(E)$ is negative. If $S (E)$ is sub-exponential, then $C (E)$ is positive, and the faster $S (E)$ grows, the larger $C(E)$ is.
Only a reservoir with infinite-dimensional Hilbert space can keep $S (E)$ growing for all $E$. And indeed, infinite-dimensional reservoirs are the ones that allow for faster cooling. However, our results are general and also apply to the finite-dimensional case.

Suppose that we want to cool a quantum system with Hilbert space dimension $d$, and Hamiltonian $H\s$ having ground-state degeneracy $g$, gap above the ground state $\Delta$, and largest energy $J$. 
What are the resources required to do so?

\medskip {\it Fundamental Assumptions.-}
Let us specify the setup more concretely and collect the assumptions we will adopt (those which come from first principles):

\noindent{\bf (i)} We consider the start of the process to be when the system has not yet been put in contact with the work storage system (the weight) nor the reservoir,
so that initially, the global state is $\rho\s \otimes \rho\b \otimes \rho_\mathsf{W}$. While other initial starting scenario may be of interest, its consideration is beyond the scope of the current paper.

\noindent{\bf (ii)} We allow for the most general quantum transformation on system, bath and weight which is reversible (unitary) and preserves total energy.
This might appear restrictive compared to paradigms which allow arbitrary interaction terms, however this is not the case, since arbitrary interactions can be incorporated into the model as  shown in Appendix H of \cite{thermoiid} and in \cite{HO-limitations}, simply by allowing the energy of the work stystem to fluctuate. In many paradigms, this is implicitly enforced by assuming that all missing energy is counted as work. 
Paradigms which relax this condition are essentially ignoring the energy transfered to other systems, or treat these other systems as classical.
Essentially, we impose energy conservation to ensure we properly account for all energy costs associated with the interaction while the various unitaries or interaction terms simply transfer or take energy from the weight to compensate. The cooling process is thus any transformation of the form
\begin{equation}
  \rho\s \ \ \to\ \  
  \mathsf{  tr}_{\sf BW}\!
  \left( U \rho\s \otimes \rho\b \otimes \rho_\mathsf{W} U^\dagger \right)\ ,
\end{equation}
where $U$ is a global unitary satisfying 
\begin{align}
  \left[ U, 
  H\s + H\b +H_\mathsf{W}
  \right] =0\ ,
\end{align}

\noindent {\bf (iii)} The work that is consumed within the transformation is taken from the weight. Since we are interested in ultimate limitations, we consider an idealized weight with Hamiltonian having continuous and unbounded spectrum $H_\mathsf{W} = \int_{-\infty}^{\infty}\! y |y\rangle\! \langle y| dy$. 
Any other work system can be simulated with this one~\cite{brandao2013second}. We denote by $\w$ the worst-case value of the work consumed, i.e.
\begin{align}
  |y-y'| >\w \ \ \Rightarrow\ \ 
  _{\sf W}\langle y| U |y'\rangle_{\sf W} =0\ .
\end{align}
$\w$ will generally be much larger than the average work $\langle W \rangle$. In any physically reasonable process carried out in finite time, one expects it to be finite.

\noindent{\bf (iv)} We also require, as in \cite{Skrzypczyk13}, that  the cooling transformation commutes with the translations on the weight. In other words, the functioning of the thermal machine is independent of the origin of energies of the weight, so that it just depends on how much work is delivered from the weight. This can be understood as defining what work is -- it is merely the change in energy we can induce on some external system.
This also ensures that the weight is only a mechanism for delivering or storing work, and is not, for instance, an entropy dump (see Result~1 in \supl). It also
ensures that the cooling process always leaves the weight in a state that can be used in the next run or the process. Thus 
\begin{align}
  \left[ U, \Pi \right] =0\ ,
\end{align}
where the Hermitian operator $\Pi$ acts as $\E^{i x\Pi} |y\rangle_{\sf W} = |y+x\rangle_{\sf W}$ for all $x,y\in \mathbb R$.
Beyond this, we allow the initial state of the weight $\rho_\mathsf{W}$ to be arbitrary. In particular, it can be coherent, which provides an advantage~\cite{thermoiid}.
%

\noindent{\bf(v)}
We assume that the bath has volume $V$ and is in the thermal state $\rho\b= \frac{1}{Z\b} \E^{-\beta H\b}$ at given inverse temperature $\beta= \frac 1 T$, with $Z\b$ the partition function of the bath. 
We denote the free energy density of the bath (in the canonical state $\rho\b$) by $f_\mathsf{can} = -\frac T {V} \ln Z\b$.

%

\noindent{\bf (vi)} The micro-canonical heat capacity~\eqref{mic HC} is not negative $C(E)$ for all energies $E$.
This implies that $S(E)$ is sublinear in $E$. We also prove in~\supl that if $S(E)$ grows linearly or faster, then perfect cooling in finite time is possible.


With these assumtions, we show that in order to perfectly cool the system to absolute zero, at least one of these two resources,
the volume of the bath $V$, or the worst-case value of the work consumed $\w$ has to be infinite.
Also, we bound the lowest achievable temperature of the system $T'\s$ in terms of $V$ and $\w$.

\medskip {\it Quantifying unattainability from first principles.-}
With assumptions (i)-(vi), we consider two cases, one where the initial and final state are thermal, and one where we allow arbitrary initial and final states. Our first result concerns the former, and states that in any process where the worst-case work injected is $\w$, the final temperature of the system cannot be lower than
\begin{eqnarray}
  \nonumber
  T\s' &\geq &
  \frac {T \Delta} 
  {V\! \left[ 
  f_\mathsf{mic}\! 
  \left(\frac {\ln \frac {2\, d} {3\, g}} {J +\frac {T\, J} {T\s} +\w}
  \right) 
  - f_\mathsf{can} \right]
  +\frac {T\, J} {T\s}
  +T\ln \frac {3 d} 
  {g}}
  \\  \label{T7-main}
\end{eqnarray}
in the large $\w, V$ limit.
The micro-canonical free energy density at inverse temperature $\beta_0$ is defined by
\begin{equation}
  \label{fmic}
  f_\mathsf{mic} (\beta_0) 
  = \frac 1 V 
  \left[ 
  E_0 - \frac 1 \beta S(E_0)
  \right]\ ,
\end{equation}
where $E_0$ is the solution of equation $  S'(E_0) = \beta_0$. 
Recall that, when the volume of the bath $V$ is large, it is usually the case that $f_\mathsf{mic} (\beta_0) = f_\mathsf{can} (\beta_0)$ and these are independent of $V$.

Let us analyze the behavior of~\eqref{T7-main} in terms of the resources invested.
As $\w$ grows, $\beta_0$ decreases and $f_\mathsf{mic}$ increases, yielding a lower final temperature $T'\s$.
Since all the volume dependence in~\eqref{T7-main} is explicit, hence, a larger $V$ also translates into a lower final temperature.

In what follows we provide a bound for the physically relevant family of entropies
\begin{equation}
  S(E) = \ln \Omega(E) = \alpha V^{1-\nu}E^\nu
  \ ,
\label{eq:radtypedos}
\end{equation}
where $\alpha>0$ and $\nu \in [1/2, 1)$ are two constants.  Such an entropy is extensive, and if we set $\nu=\frac{D}{D+1}$ it describes electromagnetic radiation (or any massless bosonic field) in a $D$-dimensional box of volume $V$.
  It is generally believed that there is no other reservoir which has a density of states growing faster with $E$ than this~\cite{unruh1982acceleration}, and certainly none which has $\nu\geq 1$. The later, corresponds to the bath with negative heat capacity discussed earlier, which enables cooling with finite $\w$. 
In the \supl\ we adapt bound~\eqref{T7-main} to the entropy~\eqref{eq:radtypedos}, obtaining
\begin{eqnarray}
  \label{Ts nu}
  T\s' \geq
  \frac {T \Delta} 
  {V\! \left[ 
  \frac {\alpha\, \nu} 
  {\ln \frac {2\, d} {3\, g}}
  (J+ \frac {T\, J} {T\s} +\w)
  \right]^{\frac 1 {1-\nu}}
  \hspace{-2mm}
  +\frac {T\, J} {T\s}
  +T\ln \frac {3 d} 
  {g}}
\end{eqnarray}
up to leading terms.
Now, all the dependence on $V$ and $\w$ is explicit. In particular, we observe that larger values of $V$ and $\w$ allow for lower temperatures. And also, larger values of $\nu$, which amount to a faster entropy growth, allowing for lower temperatures.

As mentioned above, the cooling processes that we consider are very general. In particular, they can alter the Hamiltonian of the system during the process, as long as the final Hamiltonian is identical to the initial one $H\s$. This excludes the uninteresting cooling method consisting of re-scaling the Hamiltonian $H\s \to 0$. However, our bounds can easily be adapted to process where the final Hamiltonian differs from the initial one, as we will discuss in the conclusion. 

Let us now consider the more general case, where neither the initial or final state need be thermal, but can instead be arbitrary.
As it is already well known~\cite{ticozzi2014quantum, wu2013no, reeb2013proving, brandao2013second, browne2014publisher}, the unattainability of absolute zero is not a consequence of the fact that the target state has low energy, but rather that it has low entropy. Hence, this directly translates to the unattainability of any pure state, or more generally, any state with rank $g$ lower than the initial state.
These type of processes are generally known as information erasure, or purification.
Now we analyze the limitations of any processes which takes an arbitrary initial state $\rho\s$ and transforms it into a final state $\rho'\s$ with support onto the $g$-rank projector $P$.
We quantify the inaccuracy of the transformation by the error $\epsilon = 1- \mathsf{  tr}(\rho'\s P)$.
For the sake of clarity, we assume that the system has trivial Hamiltonian $H\s =0$ (the general case is treated in the \supl), and we denote by $\lambda_{\mathsf{min}}$ and $\lambda_{\mathsf{max}}$ the smallest and largest eigenvalues of $\rho\s$.
In the \supl\ we show that any process $\rho\s \to \rho'\s$ 
has error
\begin{equation}
\label{R4-main} 
  \ln\frac 1 \epsilon 
  \leq
  \beta V\! \left[ 
  f_\mathsf{mic} \! \left(
  \frac {\ln \frac {2\, d} {3\, g}} {J +T\ln \frac {\lambda_\mathsf{max}} {\lambda_\mathsf{min}} +\w}
  \right) 
  -f_\mathsf{can} 
  \right]
  + \ln \frac 3
  {d\, \lambda_\mathsf{min}} 
\end{equation}

The results presented above, as well as others of more generality presented in the \supl, quantify our ability to cool a system (or more generally, put it into a reduced rank state), in terms of 
two resources: the volume of the bath $V$, and the worst-case fluctuation of the work consumed $\w$. They thus constitute a form of third law, in the sense that they place a bound on cooling, given some finite resources.
We now wish to translate this into the time it would take to cool the system, and we will do so, by consider the notion of a {\it Thermal Machine} and making two physically reasonable assumptions.

\medskip
{\it Thermal Machines.-}
%
%
%
%
%
%
Let us recall that the field of computational complexity is based on the Church-Turing thesis -- the idea that we consider a computer to be a Turing machine, and then explore how the time of computation scales with the size of the problem. Different machines may perform differently -- the computer head may move faster or slower across the memory tape; information may be stored in bits or in higher dimensional memory units, and the head may write 
to this memory at different speeds. Nature does not appear to impose a fundamental limit to the dimension of a computer memory unit or the speed at which it may be written. However, for any physically reasonable realisation
of a computer, and whatever the speed of these operations, it is fixed and finite, and only then do we examine the scaling of time with problem size. And what is important is the overall scaling of the time with input (polynomial or exponential), rather than any constants. Likewise here, we will consider a fixed
thermal machine, and we will assume that it can only transfer a finite amount of energy into the heat bath in finite time. Likewise, in a finite time, it cannot explore an infinite size heat bath. A thermal machine which did otherwise would be physically unreasonable.

We can consider both $V$ and $\w$ as monotonic functions of time $t$. The longer our thermal machine runs, the more work it can pump into the heat bath, and the larger the volume of the bath it can explore. For any particular thermal
machine, one can put a finite bound on $T\s'$ by substituting these functions into \eqref{Ts nu}.
In particular, if we assume that the interaction is mediated by the dynamics of a local Hamiltonian, then the interaction of a system with a bath of volume $V$ and spacial dimension $d$ will take time 
\begin{equation}
\label{t-V}
  t \geq \frac 1 v V^{1/D}\ , 
\end{equation}  
where $v$ is proportional to the speed of sound in the bath (or Lieb-Robinson velocity~\cite{LR}), and $V^{1/D}$ the linear dimension of the bath. 
The implementation of general unitaries takes much longer than~\eqref{t-V}, but this serves as a lower bound. Since we are interested here in the scaling of temperature with time, rather than with constant factors,
we need not be concerned by the fact that practical thermal machines operate at much slower speeds. 
Of course, just as with actual computers, thermal machines
generally have speeds well below the Lieb-Robinson bound.
%
Note that, despite $V$ being finite, the Hilbert space of the bath can be infinite-dimensional.  
If one wanted to have a bound which was independent of the thermal machine, and independent on the speed of sound which is a property of the bath, then one could always take $v$ to be the speed of light. While such a bound would not be practically relevant, it would be fundamental. This is similar to bounds on computation, where to get a fundamental bound, one should take the gate speed to be infinite (since there is no fundamental bound on this) and convert the number of bits used in the process to time by multiplying by the speed of light.

A relationship between worst-case work $\w$ and time $t$ is obtained by noticing the following. In finite $t$ it is not possible to inject into the bath an infinite amount of work. 
For simplicity, here we assume a linear relationship
\begin{equation}
\label{t-w}
  t \geq \frac 1 u \w\ , 
\end{equation}
where the constant $u$ will depend on the interactions between system and weight. However, we stress that, if a particular physical setup is incorrectly modeled by the relations~\eqref{t-V} and~\eqref{t-w}, then any other bound $t \geq h_1 (\w)$ and $t\geq h_2 (V)$ is also good. As long as $h_1$ and $h_2$ are strictly monotonic functions the unattainability principle will hold.

\medskip {\it Limitations using Thermal Machines.-}
%
For any particular thermal machine, we can now derive limitations on the temperature that can be reached in a given time $t$.
Since the physical system with the fastest entropy growth that we are aware of is radiation, it is worthwhile to dedicate the next paragraph to the case $\nu= \frac{D}{D+1}$ in Eq.~\eqref{eq:radtypedos}, because this should provide a bound with wide validity. Using the particular relations~\eqref{t-V} and~\eqref{t-w}, and substituting them into~\eqref{Ts nu}, for the case of radiation, we obtain
\begin{equation}
  \label{Eq9}
  T\s' \geq
  \frac {T \Delta} {v^D} 
  \left[ \frac 
  {\ln \frac {2\, d} {3\, g}} 
  {\alpha\, u} 
  \right]^{D+1}
  \hspace{-2mm}
  \frac 1 {t^{2D+1}}
  \ ,
\end{equation}
in the large $t$ limit
Our bound~\eqref{Eq9} can be straightforwardly adapted to any other relation $t \geq h_1 (\w)$ and $t\geq h_2 (V)$.
It is interesting to observe in~\eqref{Eq9} the relationship between the characteristic time (how long does it takes to cool to a fixed $T'\s$) and the size of the system $V\s$. Exploiting the usual relation $\ln d \propto V\s$ we obtain the sublinear scaling
\begin{equation}
\label{tN}
  t \propto V\s^{\frac{D+1}{2D+1}} \ .
\end{equation}



Something concerning about Result~\eqref{R4-main} is that, in the limit $\lambda_{\mathsf{min}} \to 0$ the bound becomes trivial ($\epsilon \geq 0$).
This can be solved by truncating the initial state $\rho\s$ to the subspace containing the $k$ largest eigenvalues and optimizing the resulting bound for $\epsilon$ as a function of $k$.
Also, this truncation method allows to extend all our results to infinite-dimensional systems ($d =\infty$).

\medskip {\it Discussion.-}

\begin{figure}[ht!]
\centering
\includegraphics[width=0.5\textwidth,natwidth=610,natheight=642]{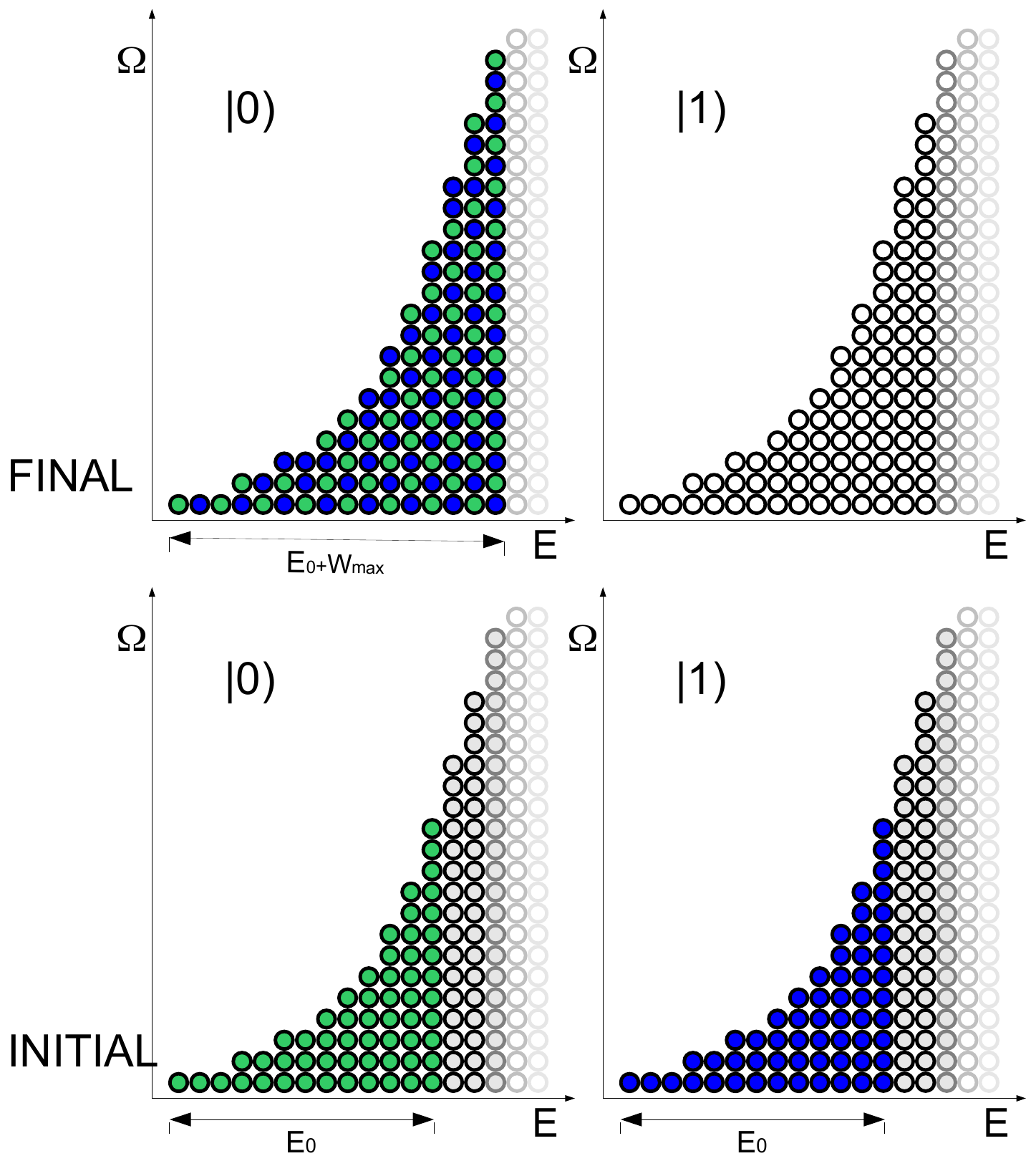}
\caption{{\bf Erasing a qubit.} Here we illustrate the limitations of transforming a qubit with $H\s =0$, from a maximally mixed state to a pure state. 
Each of the four panels depicts the function $\Omega(E)$, and each little circle represents a microstate of the bath, having the energy of the corresponding column. The two lower panels together contain all the joint states of system and bath before the transformation: the left(right) panel contains all the states of the bath together with the system being in state $|0\rangle$($|1\rangle$).
In the same way, the two upper panels contain all the joint states of system and bath after the transformation. 
The goal of erasure is to put all the states of the two lower panels to the upper-left panel, with the constraint that any state can only be shifted to the right by no more than $\w$. 
Energy $E_0$, the solution of Equation~\eqref{simp}, is the threshold below which all states from the lower two panels can be mapped to the upper-left panel. 
Above $E_0$, some states will have to be mapped to the upper-right panel, contributing to a non-zero $\epsilon$. 
}
\label{fig:holesi}
\end{figure}

The intuition behind the above results becomes more clear when we analyze a simple case.
Consider the transformation of a qubit, from a maximally mixed state to a pure state: 
\begin{equation}
\label{ppp}
  \frac 1 2 \left( |0\rangle\! \langle 0| 
  +|1\rangle\! \langle 1| \right) 
  \ \to\  
  |0\rangle\! \langle 0| \ .
\end{equation}
For simplicity we consider $H\s =0$, hence this is not really cooling, but rather erasure -- however, the essentials are identical. 
The initial(final) joint microstates of qubit and bath are depicted in the lower(upper) panels of Figure~\ref{fig:holesi}.
A perfect implementation of transformation~\eqref{ppp} amounts to mapping all the states of the lower panels to the upper-left panel.
If the Hilbert space of the bath is finite, such a transformation is incompatible with unitarity, which requires that all final states are occupied. However, an infinite-dimensional bath allows for some final states to not be the image of any initial state.
The crucial constraint is that the work that this transformation can consume in the worst case, is bounded by $\w$. This restricts the map of every state depicted in the lower panels in Figure~\ref{fig:holesi}, to a state in the upper panels which cannot be shifted to the right by more than $\w$. 
We are interested in the energy $E_0$, the threshold energy below which all initial states 
(depicted in the lower two panels) can be mapped to final states (the upper-left panel). $E_0$ is the solution of
\begin{equation}
\label{simp}
  2 I(E_0) = I(E_0 +\w)\ ,
\end{equation}
where $I(E)$ is the number of eigenvalues of $H\b$ smaller than $E$. 
The factor of two in Equation \eqref{simp} reflects the fact that there ar two initial states of the system, being mapped to one final state of the system.
The colored area of the two lower panels represent the states counted by the left-hand side of~\eqref{simp}, while the colored area of the upper-left panel represents the states counted by the right-hand side of~\eqref{simp}.
Above $E_0$, some states will have to be mapped to the upper-right panel. The error $\epsilon$ is the sum of the probabilities of all states mapped to the upper-right panel. 
In this simple case, the probability of a state from the lower panels with energy $E$ is $\frac 1 {2 Z\b} \e^{-\beta E}$ (the factor $\frac 1 2$ comes from the qubit), which decreases exponentially as we move to the right.
The optimal protocol is that which minimizes $\epsilon$. The \supl\ contains a proof that the optimal $\epsilon$ satisfies bound~\eqref{R4-main}. 
This picture makes clear that the faster $\Omega(E)$ grows, the larger $E_0$ is, and the smaller $\epsilon$ is.
But when $\Omega(E)$ grows exponentially, we have $I(E) \propto \e^{\alpha E}$, and equation~\eqref{simp} becomes $2= \e^{\alpha \w}$, which holds for a finite $\w$ independently of $E_0$.
This implies that all states can be mapped to the upper-left, and consequently $\epsilon =0$. In this and also the super-exponential case, there is no third law.

Another way to see the link between $\w$ and the achievable temperature $T'\s$  can be seen from the following protocol, recently explored, for example, in \cite{browne2013guaranteed}. 
Imagine we have a two level system with energy gap $\Delta$. One cooling method would be to put the system
in contact with a bath at temperature $T$, and raise the energy of the excited state by an amount $\w$ isothermally. 
After this, the probability that the system has successfully been put in its ground state is given by $1/(1+e^{-\beta (\Delta+\w)})$. We then
remove the system from the heat bath, and then lower the energy of the excited state back to $\Delta$, putting the system into a temperature of 
$T'_S=T\frac{\Delta}{\Delta+\w}$.   By choosing $\w$ large enough, we can make the temperature arbitrarily low, and in this limit, 
the average work $W$ is finite and dominated by the first isothermal step, giving $W=T\log{(1+e^{-\beta \Delta})}$, the change in free energy. Note that the average work $W$ is much smaller than the worst-case work $\w$.
Achieving absolute zero using this protocol clearly requires infinite resources (in this case, raising the excited state an infinite amount). 
Assumption~\eqref{t-w} applied to the above protocol gives us a final temperature
\begin{align}
T'_S(t)=T\frac{\Delta}{\Delta+u\, t}\ .
\end{align}

\medskip {\it Conclusions.-}
We hope the present work puts the third law on a footing more in line with those of the other laws of thermodynamics. These have already been long established, although they've recently been reformulated within the context of other resource theories~\cite{Werner1989, BDSW1996, uniqueinfo, thermo-ent2002, Beth-thermo, thermoiid,HO-limitations}. Namely, as described in~\cite{brandao2013second}, the first law (energy conservation), and unitarity (or microscopic reversibility) describe the class of operations which are allowed within thermodynamics. 
The zeroth law, is the fact that the only state which one can add to the theory without making it trivial, are the equivalence class of thermal states at temperature $T$. 
This allows the temperature to emerge naturally. 
The second law(s), tells us which state transformations are allowed under the class of operations. For macroscopic systems with short-range interactions, there is only one function, the entropy, which tells you whether you can go from one state to another, but in general there are many constraints~\cite{HO-limitations, brandao2013second, cwiklinski2014limitations, lostaglio2014thermodynamic}. The third law quantifies how long it takes to cool a system.  We propose to generalise it further:  While the second laws tell us which thermodynamical transitons are possible, {\it generalised third laws} quantify the time of these transitions.  In this context, it would be interesting to explore the time and resource costs
of other thermodynamical transitions. It would also be interesting to explore the third law in more restricted physical settings, as well as to other resource theory frameworks, in particular, those discussed in~\cite {brandao2013second}.

It is worth noting that scaling we find, for example, the inverse polynomial of Equation \eqref{eq:inversepoly}, is more benign than what one might have feared, and does not exclude obtaining lower temperatures with a modest increase of resources. However, it is also stronger than that envisioned when the third law was original
formulated. Consider for example, the cooling protocol of Figure~\ref{fig:wiki} proposed by Nernst. It is unphysical, since it requires
an infinite number of heat baths, each one at a lower and lower temperature, and with the final heat baths at close to zero temperature. However, it allows the temperature of the system to decrease exponentially in the number of steps, something which we are able to rule out when one doesn't have colder and colder reservoirs.

Finally, let us return to the question of the relationship between the Unattainability Principle and Nernst's Heat Theorem. In modern terms, the latter can just be understood as saying that the degeneracy of the ground state $g$ cannot be changed. This may be partly a matter of definition: if we can change the Hamiltonian then the degeneracy of the ground state can change, although one may then argue that the system is now in a different phase of matter. Regardless, one can still apply our results to this case. As described in Section~\ref{Hchange} of the \supl, this is accomplished by letting $g$ and $\Delta$ in al the above results correspond to the final Hamiltonian, while the other parameters ($J$ and $Z\s$) correspond to the initial Hamiltonian $H\s$. Thus, if the Heat Theorem is violated and we can change $g$, this at best allows us to cool at a faster rate, rather than violate the Unattainability Principle. 

\medskip
{\bf Acknowledgements:} We are grateful for discussions with Jacob Bekenstein, Fernando Brandao, Karen Hovhannisyan, Ronnie Kosloff, Micha\l\ Horodecki, Pawel Horodecki, and Mischa Woods. LM is supported by the EPSRC, and JO is supported by EPSRC and by the Royal Society.

\bibliographystyle{apsrev4-1}
%

\newpage
\section*{\supl}

In this \supl\ we detail the proofs of the results presented in the main section, and also show new results. 
In Section~\ref{sec:prelim} we we will describe the physical setup, formulate the assumptions (i)-(vi) and those given in Equations~\eqref{t-V} and~\eqref{t-w}, and introduce useful notation. 
In Section~\ref{sec:summary} we collect the statements of the key results, 
and prove them in Section~\ref{sec:proofs}. 

\section{Preliminaries}
\label{sec:prelim}

\subsection{The system being cooled}

We refer to the system to be cooled as ``the system". We assume that the system is finite-dimensional, and treat the inifinite-dimensional case in Section~\ref{id}. We denote its energy eigenstates and eigenvalues by $|s\rangle$ and $\mathcal E_s$, where $s=1,2,\ldots, d$, such that the Hamiltonian is
\begin{equation}
  H\s = \sum_s \mathcal E_s 
  |s\rangle\! \langle s| \ .
\end{equation} 
Without loss of generality $0= \min_s\mathcal E_s$ and define $J= \max_s\mathcal E_s$. We denote by $g$ the degeneracy of the ground state, and by $\Delta$ the energy gap above it. 

The initial state of the system $\rho\s$ is arbitrary, with spectral decomposition
\begin{equation}
  \rho\s = \sum_s \lambda_s
  |\phi_s \rangle\! \langle\phi_s|\ .
\end{equation}
Although special attention is given to the thermal (or Gibbs) state
\begin{equation}
  \rho\s = \frac 1 {Z\s} 
  \E^{- \beta\s H\s}
\end{equation}
at temperature $T\s = 1/\beta\s$. As usual, the normalization factor 
\begin{equation}
  Z\s = 
  \mathsf{  tr}\s\, \E^{- \beta\s H\s}
\end{equation}
is the partition function.
If the final state is thermal, we denote its temperature and partition function by $T\s'$ and $Z'\s$.

\subsection{The thermal bath}

In order to assist the cooling process there is a bath or reserboir with finite volume $V$, and hence, finite heat capacity $C$. 
The reason for limiting the volume is that, in finite time, the system can only (fully) interact with a bath of finite volume. 
At the very least, the Lieb-Robinson bound~\cite{LR} establishes a limit on the speed at which information propagates within a system with local interactions. Roughly, the time it takes for a system to interact with a $V$-volume bath, in a $D$-dimensional space, is $t \geq \frac 1 v V^{1/D}$, where $v$ is proportional to the ``speed of sound'' of the bath. 

If the volume of the bath is not well defined, then the parameter $V$ can be understood as the number of bosonic or fermionic modes, spins, subsystems, etc. Note that, despite its finite extension, the bath can have an infinite-dimensional Hilbert space. 
Even more, the bath can be a quantum field, in which case it will have an infinite number of modes. Despite this, the spatial finiteness will warrant that the number of levels with energy below any finite value will remain finite. In the pathological case where there are bosonic modes with zero energy, 
we ignore them without affecting the cooling properties of the bath.
We show below that baths with infinite-dimensional  Hilbert space have more cooling capacity.  

The energy eigenstates of the bath $|b\rangle$ are labeled by $b=1,2,\ldots$, and the corresponding energies by $\mathcal E_b$. We assume that $\mathcal E_b$ are increasingly ordered, and that $\mathcal E_1 = 0$. 
We assume that the bath is in the thermal or canonical state $\rho\b = \frac 1 {Z\b} \E^{-\beta H\b}$, at temperature $T = \frac 1 \beta$, with the normalization factor $Z\b = \mathsf{  tr}\, \E^{- \beta H\b}$ being the partition function.
The free energy of the canonical state at inverse temperature $\beta$ is
\begin{equation}
  F_\mathsf{can} = 
  -\frac 1 \beta \ln Z\b(\beta) \ ,
\end{equation}
and the density of free energy is 
\begin{equation}
\label{fed can}
   f_\mathsf{can} 
   = 
   -\frac 1 {\beta\, V} \ln Z\b (\beta)
   \ .
\end{equation}
%
The heat capacity of the canonical state is
\begin{eqnarray}
  \nonumber
  C_\mathsf{can} &=& 
  \frac {\partial \langle E \rangle} {\partial T}
  \\ \nonumber &=&
  \sum_b \frac {\mathcal E_b\, \E^{- \beta \mathcal E_b}} {Z\b} 
  \left[ -\mathcal E_b +\sum_{b'} \frac {\mathcal E_{b'}\, \E^{- \beta \mathcal E_{b'}}} {Z\b} \right]
  \frac {\partial \beta} {\partial T}
  \\ &=&
  \beta^2
  \left( \langle E^2\rangle - \langle E\rangle^2 \right)
  \ .
\end{eqnarray}

The number of states with energy at most $E$ is denoted by
\begin{equation}
  I(E) = 
  |\{b: \mathcal E_b \leq E \}|\ .
\end{equation}
The number of states with energy inside the energy window \mbox{$(E-\omega, E]$} is denoted by
\begin{equation}
\label{def Om}
  \Omega(E) = I(E)-I(E-\omega)\ ,
\end{equation}
and we refer to it as ``the density of states". 
As usual, we fix the width of the energy window matching the width of the energy distribution in the canonical state
\begin{eqnarray}
  \label{omega Ccan}
  \omega = 
  \sqrt{\langle E^2 \rangle -\langle E\rangle^2}
  = \frac {\sqrt{C_\mathsf{can}}} \beta\ .
\end{eqnarray}
It is often the case that, for large volume $V$, almost all energy levels in the interval \mbox{$(E-\omega, E]$} are clustered around the upper limit of the interval $E$. This is due to the fast increase of the number of levels as the energy grows (see for example~\cite{MMM} for a proof). When this is the case, the quantity $\Omega(E)$ does not depend much on $\omega$.

The logarithm of the density of states is the micro-canonical entropy at energy $E$
\begin{equation}
  S(E) = \ln \Omega(E)\ .
\end{equation}
It is often the case that, in the large volume limit, micro-canonical and canonical entropies become equal. This is proven in~\cite{MMM} for the case of local Hamiltonians. 
The function $S(E)$ is discontinuous, but it is usually the case that, for sufficiently large $V$, the relative size of its discontinuities is very small, and the quotient $\frac {S(E)} V$ tends to a smooth function of the energy density $E/V$ (see~\cite{MMM}).
However, here we do not make any smoothness assumption, and hence, use the discrete derivatives
\begin{eqnarray}
  \label{def S'}
  S'(E) &=& 
  \frac {S(E)-S(E-\omega)} \omega\ ,
  \\
  \label{def S''}
  S''(E) &=& 
  \frac {S(E)+S(E-2\omega) -2S(E-2\omega)}
  {\omega^2}\ .
\end{eqnarray}

The temperature of the micro-canonical state with energy $E$ is given by
\begin{eqnarray}
  T_\mathsf{mic} (E) =
  \frac 1 {S'(E)}\ ,
\end{eqnarray}
and the corresponding heat capacity is
\begin{equation}
  \label{S'' Cmic}
  C_\mathsf{mic} (E) = 
  \frac 1 {T'_\mathsf{mic} (E)} 
  =
  -\frac {S'^2(E)} {S''(E)}
  \in [0,\infty)\ .
\end{equation}
All reasonable systems (black holes excluded) have a positive heat capacity, and a negative one would allow to cool to absolute zero any system, violating the third law.
The finiteness of the heat capacity is a consequence of the finite volume of the bath.

The density of free energy for the micro-canonical state at inverse temperature $\beta_0$ is 
\begin{eqnarray}
  \label{fed mic}
  f_\mathsf{mic} (\beta_0)
  = \frac 1 V \left(
  E_0 - \frac 1 {\beta} S(E_0) 
  \right)\ ,
\end{eqnarray}
where $E_0$ is the solution of $S'(E_0) = \beta_0$.
Note the difference between the background inverse temperature $\beta$ and the one defining the state $\beta_0$.

\subsection{The cooling process}
\label{pc}

The cooling process consists of a joint transformation of system, bath and weight. We follow~\cite{Skrzypczyk13} in that the work storage device is modelled by a weight with Hamiltonian
\begin{equation}
  H_\mathsf{W} = 
  \int_{-\infty}^\infty
  \!\!\! w\, 
  |w\rangle\! \langle w|\, dw
  \ ,
\end{equation}
where the orthonormal basis $\{ |w\rangle: w \in \mathbb R \}$ corresponds to the position of the weight. 
The Hermitian operator $\Pi$ translates the energy eigenbasis
\begin{equation}
  \E^{ix \Pi}
  |w\rangle_\mathsf{W} = 
  |w+x\rangle_\mathsf{W}
  \ ,
\end{equation}
for all $w,x\in \mathbb R$.

In recent approaches to nano-thermodynamics~\cite{workvalue,HO-limitations,aaberg-singleshot}, the work consumed or generated in a process is constrained to have small fluctuations around its mean value.
However, here we want to be maximally general, hence we allow the work consumed to fluctuate arbitrarily.
Hence, our weight is an ideal source or sink of work, which can be used to simulate any other work storage device appearing in the literature. 

As mentioned in the main text, to transform cooling bounds in terms of work to one in terms of time, we can assume that injecting work into the system or the bath requires an amount of time that grows with the amount of work injected. 
Hence, since we require the transformation to be implemented within a certain given time, we restrict the worst-case work consumed to be at most a given constant $\w$. 
We stress that the work average is not restricted, and in particular, it can be larger than the free energy difference of the transformation, which breaks thermodynamical reversibility.

Abstractly, a cooling process is characterized by a completely positive and trace-preserving map
\begin{equation}
\label{GP0}
  \Gamma\s(\rho\s) = 
  \mathsf{  tr}_{\mathsf{BW}} \,
  \Gamma\!\left( 
  \rho\s \otimes \rho\b \otimes \rho_{\mathsf{W}}
  \right) 
  \ ,
\end{equation}
satisfying the following requirements:
\begin{enumerate}

  \item {\bf Microscopic reversibility:} there is a unitary operator 
\begin{equation}
  U :
  \mathcal{H\s \otimes H\b \otimes H_\mathsf{W}} \longrightarrow
  \mathcal{H\s \otimes H\b \otimes H_\mathsf{W}}\ ,
\end{equation}
jointly acting on the Hilbert spaces of system, bath and weight. That is
\begin{equation}
  \Gamma(\rho\s \otimes \rho\b \otimes \rho_{\mathsf{W}}) = 
  U \left(\rho\s \otimes \rho\b \otimes \rho_{\mathsf{W}} \right) U^\dagger\ .
\end{equation}

  \item {\bf Conservation of total energy:}
  \begin{equation}
  \label{CE}
     \big[ U,
     H\s +H\b +H_\mathsf{W}
     \big]=0
     \ .
  \end{equation}

  \item {\bf Independence of the weight's ``position":}
\begin{equation}
  \label{commut trans}
  [U, \Pi] = 0 \ .
\end{equation}


  \item {\bf Bound on worst-case work transferred:}
  \begin{eqnarray}
  \nonumber
  && U : 
  \mathcal{H\s \otimes H\b} 
  \otimes |w\rangle_\mathsf{W}       
  \longrightarrow
  \\ \label{QE1} &&
  \mathcal{H\s \otimes H\b}
  \otimes \mathsf{  span} \big\{ 
  |w'\rangle_\mathsf{W} : 
  |w' -w|\leq \w 
  \big\}
  \end{eqnarray}
  for all $w\in \mathbb R$.
\end{enumerate}
The first point follows from the fact that closed systems evolve according to the Schr\"odinger Equation. 
The second point expresses the conservation of energy (First Law of Thermodynamics). As noted in the main section, this is not a restrictive assumption, and just ensures that we account for all sources of energy. 

The third point ensures that the weight is only used as a source of work, and not, for instance, as an entropy dump~\cite{Skrzypczyk13}. More concretely, in Section~\ref{2l} it is proven that, any $U$ satisfying~\eqref{commut trans} induces a map on system and bath which is a mixture of unitaries:
\begin{equation}
  \mathsf{  tr}_{\mathsf W} \!\left(
  U \rho_{\mathsf SB} \otimes 
  \rho_{\mathsf{W}} 
  U^\dagger\right)
  = 
  \int\! dw\, p(w)\, 
  u_w \rho_{\mathsf SB} u_w^\dagger\ .
\end{equation}
This prevents the entropy of system and bath to decrease (Second Law of Thermodynamics). Essentially, for the work system, all that should matter is how much work is delivered to our system, thus
energy differences matter, but the zero of the energy should not.

The energy that is subtracted from or added to the weight is in general not fixed, it fluctuates depending on the micro-state of system and bath.
However, we assume that in finite time, the worst-case work fluctuation is bounded by a given value $\w$. 
This is encapsulated in point four. We assume that the variable $\w$ can increase with the time invested in the transformation.

\section{Summary of the argument}
\label{sec:summary}

In this section we summarize the derivation of the quantitative third law by presenting the steps of the proof as a series of results. The proofs are given in Section \ref{sec:proofs}. 

We start by showing that the action of $U$ on system and bath (when the weight is traced out) cannot reduce their entropy. In particular, this prevents to use the weight as an entropy dump. 
\begin{result}\label{2law}
If $[U, \Pi]=0$ then for every state of the weight $\rho_{\mathsf W}$ there is a probability distribution $p(w)$ such that
\begin{eqnarray}
  \nonumber
  \Gamma_{\sf SB} (\rho_{\sf SB})
  &=& 
  \mathsf{  tr}_{\mathsf W} \!\left(
  U \rho_{\mathsf SB} \otimes 
  \rho_{\mathsf{W}} 
  U^\dagger\right)
  \\ &=&
  \int\! dw\, p(w)\, 
  u_w \rho_{\sf SB} u_w^\dagger\ ,
\end{eqnarray}
where $u_w$ are unitaries.
\end{result}
As a result, the action of the cooling process on the system and bath is at best unitary (which will preserver the Von Neumann entropy), and possibly entropy increasing if a mixture of unitaries.
As we will show in Section~\ref{2l}, the unitaries $u_\omega$ depend on the global unitary $U$, but not on the state of the weight $\rho_{\mathsf W}$.

The error of the cooling process is quantified by the probability of not being in the ground space
\begin{equation}
  \epsilon = 
  1- \mathsf{  tr}
  [P\, \Gamma\s (\rho\s) ]
  \ ,
\end{equation}
where $P$ is the projector onto the ground space of $H\s$ and $\Gamma\s$ is the action of the cooling process on the system
\begin{equation}
  \Gamma_{\sf S} (\rho_{\sf S})
  = 
  \mathsf{  tr}_{\mathsf BW} \!\left(
  U \rho\s \otimes \rho\b\otimes 
  \rho_{\mathsf{W}} 
  U^\dagger\right)
  \ .
\end{equation}
For the following result it is useful to define 
\begin{equation}
\label{xi 34}
  \xi = J +T\ln\frac {\lambda_{\sf max}} {\lambda_{\sf min}} +\w
  \ ,
\end{equation}
where $\lambda_{\sf max}, \lambda_{\sf min}$ are the largest and smallest eigenvalues of $\rho\s$.

\begin{result}\label{RLT0}
The error made when cooling a system to absolute zero
satisfies
\begin{equation}
  \label{eps R0}
  \epsilon \geq
  \frac {\E^{-\beta (E_0 +\omega)}} {Z\b} \lambda_{\mathsf{min}}
  \left[
  d \Omega (E_0 +\omega)
  -g \Omega (E_0+ \xi +\omega)
  \right]
\end{equation}
where $E_0$ is the smallest energy violating
\begin{equation}
\label{R balanc2}
  d I (E) \leq
  g\, I(E +\xi)\ .
\end{equation}
\end{result}
Recall that $\omega$ is defined in~\eqref{omega Ccan}.
The dependence of~\eqref{eps R0} and~\eqref{xi 34} on the smallest eigenvalue $\lambda_{\sf min}$ makes $\epsilon$ a discontinuous function of the state $\rho\s$, which is unphysical. In Section~\ref{id} we apply a standard smoothing technique to make $\epsilon$ a continuous function of the state. This also allows us to adapt our results to infinite-dimensional systems $(d \to \infty)$.

The above bound on $\epsilon$ is valid with full generality. However, solving equation~\eqref{R balanc2} is in general difficult.
Next we assume the positivity of the (micro-canonical) heat capacity and derive a more usable bound.
\begin{result}\label{R S'>0}
The error made when cooling a system to absolute zero satisfies
\begin{equation}
  \label{eps bound simple}
  \epsilon 
  \geq
  \frac 1 {Z\b} 
  \E^{-\beta (E_0+\omega) +S(E_0+\omega)} 
  \, \lambda_{\mathsf{min}}
  \, \frac {d} 3 \ ,
\end{equation}
where $E_0$ is the (unique) solution of
\begin{equation}
  \label{S' E0 xi}
  S'(E_0)\, \xi = \ln \frac {2\, d} {3\, g}\ ,
\end{equation}
provided 
\begin{equation}
  \label{premise}
  \frac T {\sqrt 2} \,
  C^{1/2}_\mathsf{can} (V) > 
  J +T\ln\frac {\lambda_{\sf max}} {\lambda_{\sf min}} +\w
  \ ,
\end{equation}
and
\begin{equation}
  \label{premise 2}
  0\leq 
  C_\mathsf{mic} (E)
  < \infty\ , 
\end{equation}
for all $E$.
\end{result}
Our bound~\eqref{eps bound simple} depends implicitly on the two parameters $V$ and $\w$, which quantify the amount of resources.
But bound~\eqref{eps bound simple} is valid in the range of parameters ($V$ and $\w$) satisfying condition~\eqref{premise}.
However, this regime includes the late time situations we are interested in, since we can take the volume of the bath $V$ sufficiently large, and $C_\mathsf{can} (V)$ grows with the volume in at least a linear rate.

An important consequence of Result~\ref{R S'>0} is the following. The faster $\Omega (E)$ or $S(E)$ grow, the slower $S'(E)$ decreases and the larger the solution $E_0$ of equation~\eqref{S' E0 xi} is.
A large $E_0$ in~\eqref{eps bound simple} gives a smaller bound for $\epsilon$.  
As we will see in Result~\ref{RLT7}, baths with faster $\Omega$-growth allow for better cooling.
And actually, when this growth is exponential or faster then perfect cooling ($\epsilon =0$) can be achieved with finite $\w$ (see Section~\ref{hilberthotel}).
However, these cases correspond to negative heat capacity, which is unphysical.


\begin{result}\label{RLT5}
If the final state is thermal, then its temperature $T'\s$ satisfies
\begin{equation}\label{RLT5f}
  T\s' \geq
  \frac \Delta 
  {\ln \frac d {g\, \epsilon}}
  \ ,
\end{equation}
where $\epsilon$ is the probability of the system not being in the ground state~\eqref{eps bound simple}, and $\Delta$ is the energy of the first excited state above the ground space of the system.
\end{result}
The following result applies Result~\ref{R S'>0} to the case where the initial state of the system is thermal at temperature $T\s$, in which case we have
\begin{equation}
  \xi = J + \frac T {T\s} J +\w\ .
\end{equation}
It also uses Result~\ref{RLT5} to translate the error probability $\epsilon$ to the final temperature of the system $T'\s$.
%

\begin{result}\label{RLT6}
If the microcanonical entropy of the bath scales as
\begin{equation}
  S(2E,2V)=2S(E,V)
\end{equation}
then, the final temperature of the system cannot be lower than
\begin{eqnarray}
  \nonumber
  T\s' &\geq &
  \frac {T \Delta} 
  {V\! \left[ 
  f_\mathsf{mic} (\frac 1 \xi \ln \frac {2\, d} {3\, g}) 
  - f_\mathsf{can}(\beta) \right]
  +\frac T {T\s} J
  +T\ln \frac {3 d} 
  {g}}
  \\   \label{T7}
\end{eqnarray}
where only leading terms in $V$ are considered, and we take the regime given by Equations~\eqref{premise} and~\eqref{premise 2}.
\end{result}
This result appears in the main section, as Equation~\eqref{T7-main}. 
Here we can observe the very natural fact that: the smaller the initial temperature $T\s$ is, the smaller the final temperature $T'\s$ becomes. The above result is already a third law, in the sense that it places a limitation on the temperature which is achievable, given a restriction on resources $V$ and $\w$.
Also note that, in most known types of bath, when $V$ is large we have $f_\mathsf{mic}(\beta_0) \approx f_\mathsf{can}(\beta_0)$, and no distinction between the two free energy densities is necessary.

In what follows we consider a family of entropy functions for the bath. In particular, this family contains the entropy of a  box of electro-magnetic radiation in $D$ spatial dimensions and volume $V$ (in the large $V$ limit).
This result illustrates, that the faster the entropy grows (larger $\nu >0$), the lower the achievable temperature.

\begin{result}\label{RLT7}
If the entropy function of the bath is 
\begin{equation}
  \label{SVE}
  S(E) = \alpha V^{1-\nu} E^{\nu}
  \ ,
\end{equation}
with $\nu \in [1/2, 1)$, then equation~\eqref{T7} becomes
\begin{eqnarray}
  T\s' &\geq &
  \frac {T \Delta} 
  {V\left[ 
  \frac {\alpha} 
  {\ln \frac {2\, d} {3\, g}}\, \xi
  \right]^{\frac 1 {1-\nu}}
  \hspace{-2mm}
  +\frac T {T\s} J
  +T\ln \frac {3 d} 
  {g}}
  \ ,
\end{eqnarray}
up to leading terms in $V$ and $\xi$.
\end{result}

The largest work fluctuation $\w$ and the volume of the bath $V$ are resources that we associate to the time invested in the cooling process. The larger this quantities are, the lower the final temperature can be. In the following result we express the lowest achievable temperature in terms of time. In order to facilitate a simpler expression above, we have suppressed all constant terms.

\begin{result}\label{RLT8}
If the entropy function is~\eqref{SVE} and
\begin{eqnarray*}
  t &\geq & 
  \mbox{$\frac 1 v$}\, 
  V^{1/D}\ ,
\\
  t &\geq & 
  \mbox{$\frac 1 u$}\, \w\ ,
\end{eqnarray*}
then
\begin{equation}
  T\s' \geq
  \frac {T \Delta} {v^D} 
  \left[ \frac 
  {\ln \frac {2\, d} {3\, g}} 
  {\alpha\, u} 
  \right]^{D+1}
  \hspace{-2mm}
  \frac 1 {t^{2D+1}}
  \ ,
\end{equation}
up to leading terms.
\end{result}

\subsection{Cooling processes with non-constant $H\s$}
\label{Hchange}

Our previous results only apply to the case where the Hamiltonian of the system is kept constant during the cooling process. However, our results can be easily adapted to the case where the Hamiltonian changes. 

It is straightforward to check that, if the parameters $g$ and $\Delta$ appearing in our formulae are those of the final Hamiltonian, then Results~\ref{RLT0}, \ref{R S'>0}, \ref{RLT5}, \ref{RLT6}, \ref{RLT7}, \ref{RLT8} become true for processes with non-constant $H\s$.

\subsection{Cooling processes which discard part of the system}

A well known example of this type of process is evaporative cooling, in which the cooled final system contains only a fraction of the atoms of the initial system.
In general, we write the Hilbert space of the initial system $\hcal\s$ as the product of the final system $\hcal\s'$ times the discarded part $\hcal\s''$, that is $\hcal\s = \hcal\s' \otimes\hcal\s''$. This translates into a relation between the respective dimensions of these Hilbert spaces $d=d' d''$.
 
Now, we can repeat the argument that led to Results~1-7, but replacing $\hcal\s^0$ by our new target subspace $\hcal\s'^0 \otimes\hcal\s''$, where $\hcal\s'^0$ is the ground space of $\hcal\s'$. Hence, if $P'$ is the projector onto $\hcal\s'^0$ then the projector onto the target subspace is $P=P' \otimes \unity\s''$ where $\unity\s''$ is the identity of $\hcal\s''$.
Also, if $g$ and $g'$ are the ranks of $P$ and $P'$, respectively, then we have the relation $g=g'd''$.

Now we apply our results to subspace $P$.
We just have to keep in mind that $d, g, J, \lambda_{\min}, \lambda_{\max}, T\s$ refer to the initial system (before the partial trace), and $\Delta, T\s'$ refer to the final system.
Also note that the dependence of our bounds on $d$ and $g$ is via $d/g$, hence there is no difference when using either $d/g$ or $d'/g'$. 

\subsection{Continuity and infinite-dimensional systems}
\label{id}

Some of our bounds on the error $\epsilon$ and the temperature $T'\s$ depend on the lowest eigenvalue $\lambda_{\sf min}$, which makes them a discontinuous function of the initial state $\rho\s$. For example, Result~\ref{R S'>0} provides a lower bound for $\epsilon$ which tends to zero as $\lambda_{\sf min}$ tends to zero. Hence, for very small $\lambda_{\sf min}$ the result is useless.

One way to fix this problem is by truncating the Hilbert space of the system $\hcal\s$ such that the smallest eigenvalues of $H\s$ are eliminated. After truncating $\hcal\s$, system's parameters $\lambda'_{\sf min}$, $J'$ and $d' $ may take new values. 
The physical meaning of truncation is that the truncated subspace is assumed to be mapped to the ground space without interfering with the map of the untruncated subspace. Hence, the truncated subspace does not contribute to the error $\epsilon$.

Ideally, one should optimise over all possible truncations, until the lower bounds for $\epsilon$ and $T'\s$ are maximal. The truncation always has an optimal non-trivial point, because if one truncates everything except for a 1-dimensional subspace then $\epsilon=0$, which cannot be maximal.
Hence, there is an optimal truncation dimension $d' $ in the interval $1 < d'  \leq d$ for which $\epsilon$ and $T'\s$ are maximal.

This method can also be used to apply all our results to infinite-dimensional systems. Any finite-dimensional truncation gives finite $\lambda'_{\sf min}$, $J'$ and $d' $, which provide a non-trivial bound when substituted in any of our results.

Now, let us apply the truncation method to a harmonic oscillator with energy levels $\mathcal E_s = s$ for $s=1, 2, \ldots$, with initial state being a thermal state at the same temperature as the bath $T\s =T$. This system suffers from the above mentioned two problems: 
 $\lambda_{\sf min} =0$ and $d=\infty$.
We solve these problems by truncating out all energy levels $s > d'$, obtaining
\begin{eqnarray}
  J' &=& d'\ ,
  \\
  \lambda'_{\sf min} &=& 
  \frac {\e^{-\beta d'}} {Z\s}\ ,
  \\
  \xi' &=& 2 d'+\w\ .
\end{eqnarray}
Substituting this in~\eqref{eps bound simple} we obtain
\begin{eqnarray}
  \label{Eq 69}
  \epsilon 
  &\geq &
  \frac 1 {Z\b Z\s} 
  \E^{-\beta (E_0+\omega +d') +S(E_0+\omega)} 
  \, \frac {d'} 3 \ ,
\\
  S'(E_0) &=& 
  \frac {\ln \frac {2\, d'} {3}}
  {2d' +\w}\ .
\end{eqnarray}
Setting $d' =2$ gives a non-trivial bound. However, to obtain the optimal value of $d'$, one needs to jointly optimize the above two equations.

\subsection{Negative heat capacity allows perfect cooling}
\label{hilberthotel}

In this subsection we show that when the heat capacity of the bath is negative
\begin{equation}
  C_\mathsf{mic} (E)
  < 0\ ,
\end{equation}
for all $E$, then perfect cooling to absolute zero ($\epsilon =0$) with finite $\w$ is possible.
Using relation~\eqref{S'' Cmic} we see that the violation of the above conditions implies $S''(E) \geq 0$. Which in turn implies that $S(E)$ grows at least linearly, or that $\Omega (E)$ grows at least exponentially.

First note that the integral $I(E)$ never grows slower than $\Omega(E)$. Hence $I(E)$ is also exponential or super-exponential. This implies that for  any value of $d/g$ there is a sufficiently large energy $\w$ such that
\begin{equation}
  \label{ineq140}
  \frac {I (E+\w)} {I(E)}
  \geq \frac {d} g \ ,
\end{equation}
for all $E$.
Therefore the smallest energy violating~\eqref{ineq140} is $E_0 = \infty$, which, when substituted in~\eqref{eps R0}, gives $\epsilon = 0$.
In other words, the whole space $\hcal\s \otimes \hcal\b$ can be mapped into the ground space $\hcal\s^0 \otimes \hcal\b$. There is no unattainability principle.

\section{Proofs}
\label{sec:proofs}

\subsection{The weight as a work storage system (Result~\ref{2law})}
\label{2l}

In this subsection we show that, if the global transformation $U$ commutes with the translations on the weight then the effect of $U$ on system and bath is a mixture of unitaries. And hence, it can never decrease the entropy of system and bath, which amounts to a statement of the second law of thermodynamics. 

Let $U$ be a unitary acting on system, bath and weight $\mathcal H_\mathsf{SB} \otimes \mathcal H_\mathsf{W}$ which commutes with the translations on the weight $[U, \unity_{\sf SB} \otimes \Pi]=0$. This implies that we can write it as
\begin{equation}
  U= \int\! dx\, A_x 
  \otimes \E^{ix \Pi}\ ,
\end{equation}
where $A_x$ is a family of operators acting on $H_\mathsf{SB}$.
By imposing unitarity we obtain
\begin{equation}
  \unity_{\sf SB}\otimes\unity_{\sf W} 
  = UU^\dagger = 
  \int\! dx\, dx'\, 
  A_x A_{x'}^\dagger 
  \otimes \E^{i(x-x') \Pi}\ ,
\end{equation}
which implies
\begin{equation}
  \label{dI}
  \int\! dx\, A_x A_{x+y}^\dagger 
  = \unity_{\sf SB}\, \delta(y) \ ,
\end{equation}
where $\delta(y)$ is the Dirac delta distribution.
Now, let us figure out the structure of the reduced map on system and bath:
\begin{eqnarray}
  \nonumber
  \Gamma(\rho) &=& \mathsf{  tr}_\mathsf{W} \! \left(
  U \rho \otimes \sigma U^\dagger \right)
  \\ \nonumber &=&
  \int\! dx\, dx'\, A_x \rho A_{x'}^\dagger 
  \ \mathsf{  tr}_\mathsf{W} \! 
  \left( \E^{i(x-x') \Pi} \sigma \right)
  \\ &=&
  \int\! dx\, dx'\, A_x \rho A_{x'}^\dagger 
  \int\! dk\, p(k)\, e^{ik(x-x')}
\end{eqnarray}
where $p(k) = \langle k | \sigma |k\rangle$ is a probability measure and $|k\rangle$ is the eigenstate $\Pi |k\rangle = k |k\rangle$ for all $k\in \mathbb R$.
Also, note that, in general, the  map $\Gamma$ depends on the state of the weight $\sigma$.

If we define the family of operators
\begin{equation}
  u_k =   \int\! dx\, A_x\, e^{ikx}\ , 
\end{equation}
then we can write the reduced map as
\begin{equation}
  \label {Rm}
  \Gamma(\rho) = 
  \int\! dk\, p(k)\, 
  u_k \rho\, u_k^\dagger\ .
\end{equation}
Now we show that the operators $u_k$ are unitary:
\begin{eqnarray}
  \nonumber
  u_k u_k^\dagger &=& 
  \int\! dx\, dx'\, A_x A_{x'}^\dagger 
  e^{ik(x-x')}
  \\ \nonumber &=&
  \int\! dx\, dy\, A_x A_{x+y}^\dagger 
  e^{-iky}
  \\ &=&
  \int\! dy\, \delta(y)\, 
  \unity\, e^{-iky}
  = \unity\ ,
\end{eqnarray}
where we have used~(\ref{dI}).
In summary, for any initial state of the weight $\sigma$, the reduced map~\eqref{Rm} is a mixture of unitaries. Interestingly, the set of unitaries $u_k$ is independent of $\sigma$, but the probability measure $p(k)$ does depend on $\sigma$.

\subsection{Optimal cooling (Result~\ref{RLT0})}
\label{SB}

Here, we formalize the intuition described in Figure~\ref{fig:holesi} for quantifying the probability of all the states which cannot be mapped to the ground space.

We define the subspaces with energy lower than a given value $E$ as
\begin{eqnarray}
  \hcal \s ^E &=&
  \mathsf{  span}\! \left\{ 
  |s\rangle: \mathcal E_s \leq E
  \right\} \subseteq \hcal\s\ ,
  \\
  \hcal \b ^E &=&
  \mathsf{  span}\! \left\{ 
  |b\rangle: \mathcal E_b \leq E
  \right\} \subseteq \hcal\b\ ,
  \\
  \hcal _{\mathsf SB} ^E &=&
  \mathsf{  span}\! \left\{ 
  |s\rangle |b\rangle: \mathcal E_s +\mathcal E_b \leq E
  \right\} \subseteq 
  \hcal\s \otimes \hcal\b\ ,\quad
\end{eqnarray}
and recall that
\begin{eqnarray}
  \mathsf{  dim}\, \hcal \s ^0 &=& g\ ,
  \\ \label {IB}
  \mathsf{  dim}\, \hcal \b ^E &=& I(E)\ .
\end{eqnarray}
The optimal cooling unitary $u$ is the one that maps the largest amount of probability from $\rho\s \otimes \rho\b$ to the ground space $\hcal\s^0 \otimes \hcal\b$. 
Hence, it is useful to denote the subspace corresponding to the largest eigenvalues of $\rho\s \otimes \rho\b$ by
\begin{eqnarray}
  \nonumber
  \mathcal Q_X &=& 
  \mathsf{  span}\!
  \left\{ |\phi_s\rangle |b\rangle : 
  \lambda_s\, p_b \geq 
  \frac {\E^{-\beta X}} {Z\b}
  \right\}
  \\ \label{QE} &=&
  \mathsf{  span}
  \big\{ |\phi_s \rangle |b\rangle : 
  \mathcal E_b \leq X +T \ln \lambda_s
  \big\}
  \\ \nonumber &\subseteq &
  \mathcal {H\s \otimes H\b}\ ,
\end{eqnarray}
where $X$ is a convenient way to parametrize the probability, and we have used $p_b = \frac 1 {Z\b} \E^{-\beta \mathcal E_b}$.
The dimension of this ``large probability" subspace is
\begin{eqnarray}
  \nonumber
  \mathsf{  dim}\, 
  \mathcal Q_X  
  &=& \sum_s I\!\left(
  X +T\ln \lambda_s \right)
  \\ \label{dimQ} &\geq &
  d\, I\!\left(
  X +T\ln \lambda_\mathsf{min} \right)\ ,
\end{eqnarray}
where $\lambda_\mathsf{min} = \min_s \lambda_s$.
Using $J= \max_s \mathcal E_s$ and~\eqref{QE} we obtain
\begin{equation}
  \mathcal Q_X  \subseteq
  \hcal _{\mathsf SB} ^{X+ J+ T\ln \lambda_{\sf max}} \ ,
\end{equation}
where $\lambda_\mathsf{max} = \max_s \lambda_s$.
Assumption ``Bound on worst-case work transferred", stated in~\eqref{QE1}, can be written as
\begin{equation}
  \label{wwwww}
  u (\hcal _{\mathsf SB} ^{E})
  \subseteq
  \hcal _{\mathsf SB} ^{E+ \w}\ .
\end{equation}
Combining the two above equations we obtain
\begin{equation}
  \label{XSB}
  u (\mathcal Q_X) 
  \subseteq 
  \hcal_{\mathsf SB}^{X+ w_0} \ ,
\end{equation}
where
\begin{equation}
  w_0 = 
  J+T \ln \lambda_{\sf max} +\w\ .
\end{equation}
Note that, if $\mathcal Q_X$ is (fully) mapped into the ground space $\hcal\s^0 \otimes \hcal\b$ then~\eqref{XSB} implies
\begin{equation}
  \label{relevant 86}
  u (\mathcal Q_X)
  \subseteq 
  \hcal\s^0 \otimes
  \hcal_{\sf B}^{X+ w_0} \ ,
\end{equation}
which in turn implies (substituting~\eqref{dimQ} and~\eqref{IB})
\begin{equation}
\label{balanc}
  d I( X +T\ln \lambda_{\sf min})
  \leq
  g\, I(X+ w_0)\ .
\end{equation}
For the subsequent analysis it is convenient to define the threshold value $X_0$, which is the infimum of the $X$'s violating~\eqref{balanc}.
With this definition, we write the decomposition
\begin{equation}
  \hcal\s \otimes \hcal\b  
  = Q _{X_0 +\omega} \oplus 
  Q _{X_0 +\omega}^\perp
\end{equation}
in orthogonal subspaces.

In order to obtain an upper bound for the amount of probability that can be mapped from $\rho\s \otimes \rho\b$ to the ground space, we assume that there is no constraint on where $\mathcal Q _{X_0 +\omega} ^\perp$ is mapped, 
and that the only constraint on the image of $\mathcal Q _{X_0 +\omega}$ is~\eqref{XSB}. 
However, the definition of $X_0$ vie~\eqref{balanc} prevents mapping all of $\mathcal Q _{X_0 +\omega}$ into the ground space.
Clearly, the optimum is to map the subspace of $\mathcal Q _{X_0 +\omega}$ containing the largest eigenvalues of $\rho\s \otimes \rho\b$ to the ground space. The complement of this subspace
\begin{equation}
  \mathcal W = 
  \mathcal Q _{X_0 +\omega}
  \ominus u^{-1}\! 
  \left(\hcal\s^0 \otimes
  \hcal\b ^{X_0+ w_0 +\omega}\right)
  \ ,
\end{equation}
cannot be mapped into the ground space.
Also, we know it has dimension
\begin{eqnarray*}
  \mathsf{  dim} \mathcal W 
  &\geq & 
  d I (X_0 +T\ln \lambda_\mathsf{min} +\omega)
  -g I (X_0+ w_0 +\omega)
  \\ &= &
  d \Omega (X_0 +T\ln \lambda_\mathsf{min} +\omega)
  -g \Omega (X_0+ w_0 +\omega)
  \\ &&+
  d I (X_0 +T\ln \lambda_\mathsf{min})
  -g I (X_0+ w_0)
  \\ &\geq &
  d \Omega (X_0 +T\ln \lambda_\mathsf{min} +\omega)
  -g \Omega (X_0+ w_0 +\omega)
\end{eqnarray*}
where in the last inequality we have used the definition of $X_0$ via~\eqref{balanc}.
The subspace $\mathcal W$ is mapped in the complement of the ground space, and hence, it contributes to the error $\epsilon$. Our lower-bound for $\epsilon$ is obtained by only taking into account this contribution.

Equation~\eqref{QE} tells us that the smallest eigenvalue in $\mathcal Q _{X_0 +\omega}$, and hence in $\mathcal W$, is not smaller than $\frac 1 {Z\b} \E^{-\beta (X_0+\omega)}$.
Therefore, we can bound $\epsilon$ by the product of this number with the dimension of $\mathcal W$
\begin{eqnarray}
  \label{eps 88}
  \epsilon &\geq &
  \frac {\E^{-\beta (X_0 +\omega)}} {Z\b}
  \\ \nonumber && \times
  \left[
  d \Omega (X_0 +T\ln \lambda_\mathsf{min} +\omega)
  -g \Omega (X_0+ w_0 +\omega)
  \right]
  \\ \nonumber &= &
  \frac {\E^{-\beta (E_0 +\omega)}} {Z\b} \lambda_{\mathsf{min}}
  \left[
  d \Omega (E_0 +\omega)
  -g \Omega (E_0+ \xi +\omega)
  \right]
\end{eqnarray}
where we have used definitions
\begin{eqnarray}
  \label {def E_0}
  E_0 &=& X_0+ T\ln \lambda_{\mathsf{min}}\ ,
  \\
  \xi &=& J +T\ln\frac {\lambda_{\sf max}} {\lambda_{\sf min}} +\w\ .
\end{eqnarray}
In these new variables, $E=E_0$ is the smallest energy violating
\begin{equation}
  \label{balanc E}
  d I(E) \leq g I(E+\xi)\ .
\end{equation}

\subsection{Simpler bound for the error (Result~\ref{R S'>0})}

In this subsection we derive an upper bound
\begin{equation}
  E_1 \geq E_0\ ,
\end{equation}
which is easier to obtain than solving~\eqref{balanc E}.
This bound $E_1$ is used to write a lower bound for $\epsilon$ that is simpler than~\eqref{eps 88}.
We start by assuming that $E_1$ satisfies 
\begin{equation}
  \label{assump S'E_1}
  S'(E_1) >
  \frac {1.3\, \beta}
  {\sqrt{C_\mathsf{can}}}\ ,
\end{equation}
and later we prove that this  assumption reduces to premise~\eqref{premise}. 
On the other hand, premise~\eqref{premise 2} warrants that
\begin{equation}
  \label{S''<0}
  S''(E) \leq 0\ ,
\end{equation}
which it is used below.

Taylor's theorem implies that for any pair $E,\xi >0$ there is $\xi^* \in [0,\xi]$ such that
\begin{eqnarray}
  S(E+\xi) =
  S(E) + S'(E) \xi + 
  S''(E+\xi^*) \frac {\xi^2} 2
  \ .
\end{eqnarray}
This and~\eqref{S''<0} imply
\begin{eqnarray}
  \label{S''ineq} 
  S(E+\xi) &\leq & 
  S(E) + S'(E) \xi\ .
\end{eqnarray}
Assumption~\eqref{assump S'E_1} implies $S'(E) > 0$, which together with~\eqref{S''ineq} gives the upper bound
\begin{eqnarray}
  \nonumber
  I(E+\xi) 
  &= & 
  \sum_{k=0}^\infty \Omega(E+\xi-\omega k) 
  \\ \nonumber &\leq &
  \sum_{k=0}^\infty \, 
  \e^{S(E) + S'(E) (\xi-\omega k)}
  \\ \nonumber &= &
  \e^{S(E) + S'(E)\xi}
  \sum_{k=0}^\infty \, 
  \e^{-S'(E)\omega k}
  \\ \label{upI} &= &
  \frac 
  {\e^{S(E) +S'(E)\xi}} 
  {1-\e^{-S'(E)\omega}} 
  \ .
\end{eqnarray}
We can also write the lower bound
\begin{equation}
  \label{loI}
  I(E) \geq \Omega(E) = \e^{S(E)} \ .
\end{equation}
Substituting bounds~\eqref{upI} and~\eqref{loI} in~\eqref{balanc E} we obtain 
\begin{equation}
  \label{penultimate eq E_thr}
  d
  \leq g\,
  \frac 
  {\e^{S'(E)\, \xi}} 
  {1-\e^{-S'(E)\, \omega}} \ .
\end{equation}
Substituting $\omega = \sqrt {C} / \beta$ and using assumption~\eqref{assump S'E_1} we obtain $1-\e^{-S''(E) \sqrt{C}/\beta} > \frac 2 3$, which allows us to write~\eqref{penultimate eq E_thr} as
\begin{equation}
  S'(E)\, \xi 
  \geq 
  \ln \frac {2\, d} {3\, g}\ .
\end{equation}
We define $E_1$ to be the infimum value of $E$ violating~\eqref{ultime eq E_thr}, which satisfies \jono{??}
\begin{equation}
  \label{ultime eq E_thr}
  S'(E_1)\, \xi 
  = 
  \ln \frac {2\, d} {3\, g}\ .
\end{equation}
If $S'(E)$ is strictly monotonic, then equation~\eqref{ultime eq E_thr} has a unique solution $E_1$.
Let us show that $S'(E)$ is strictly monotonic.

Equation~\eqref{ultime eq E_thr} implies $S'(E_1)>0$, which together with the finiteness of the micro-canonical heat capacity~\eqref{premise 2} forces $S''(E_1) <0$. And this in turn implies the strict decreasing monotonicity of $S'(E)$ around $E_1$.

Substituting~\eqref{ultime eq E_thr} in~\eqref{assump S'E_1} gives $\sqrt C > 1.3\, \beta\, \xi$, which is premise~\eqref{premise}. This proves our previous claim: assumption~\eqref{assump S'E_1} reduces to premise~\eqref{premise}.  

Now, substituting $E_1$ in~\eqref{eps 88} and using~\eqref{S''ineq} we obtain
\begin{eqnarray*}
  \label{eps R0 X}
  \epsilon &\geq &
  \frac {\E^{-\beta (E_1 +\omega)}} {Z\b} \lambda_{\mathsf{min}}
  \left[ d\, \e^{S(E_1+\omega)} 
  -g\, \e^{S(E_1 +\xi +\omega)}
  \right]
  \\ &\geq &
  \frac {\E^{-\beta (E_1 +\omega)}} {Z\b} 
  \lambda_{\mathsf{min}}\,
  \e^{S(E_1+\omega)} 
  \left[ d
  -g\, \e^{S'(E_1 +\omega) \xi}
  \right]\ ,
\end{eqnarray*}  
and using the monotonicity of $S'(E)$ we get
\begin{eqnarray}
  \nonumber
  \epsilon &\geq &
  \frac {\E^{-\beta (E_1 +\omega)}} {Z\b} 
  \lambda_{\mathsf{min}}\,
  \e^{S(E_1+\omega)} 
  \left[ d
  -g\, \e^{S'(E_1) \xi}
  \right]
  \\ \label{SC eps bound} &= &
  \frac 1 {Z\b} 
  \E^{-\beta (E_1+\omega) +S(E_1+\omega)} 
  \, \lambda_{\mathsf{min}}
  \, \frac {d} 3 \ ,
\end{eqnarray}
where the equality follows from~\eqref{ultime eq E_thr}.

\subsection{Relationship between error and temperature (Result~\ref{RLT5})}

In this subsection we assume that the final state is thermal at temperature $T\s'$ and has partition function $Z\s' = Z\s (T\s')$. Because of our convention $\min_s \mathcal E_s = 0$, we have
\begin{equation}\label{q-Z}
  \epsilon = 1-\frac g {Z\s'}\ .
\end{equation}
In principle, the function $Z\s (T\s')$ can be inverted, and the bound for $\epsilon$~\eqref{eps bound simple} can be transformed into a bound for $T\s'$.
However, this is in general a hard task.
In what follows we obtain a general relation between $Z'\s$ and $T\s'$ which avoids having to invert $Z\s (T\s')$.

For any Hamiltonian $H\s$ following convention $\min_s \mathcal E_s = 0$ we have
\begin{equation}
  Z\s ' \leq g + d \E^{-\Delta /T\s'}
  \ .
\end{equation}
Combining this with~\eqref{q-Z} we obtain
\begin{equation}
  \epsilon \leq 
  \frac {d} g \E^{-\Delta/T\s'}
  \ ,
\end{equation}
or equivalently,
\begin{equation}
\label{T'}
  T\s' \geq
  \frac \Delta 
  {\ln\frac {d} {\epsilon\,g}}
  \ .
\end{equation}

\subsection{Cooling the system from $T\s$ to $T'\s$}

In this section we assume that the initial state of the system is thermal $\lambda_s = \frac 1 {Z\s} \E^{-\mathcal E_s /T\s}$ at temperature $T\s$.
In this case we have
\begin{eqnarray}
  \lambda_{\mathsf{max}} 
  &=& \frac 1 {Z\s}\ ,
  \\
  \lambda_{\mathsf{min}} 
  &=& \frac {\E^{-J/T\s}} {Z\s}\ ,
  \\
  \xi &=& J + \frac {T}{T\s} J +\w \ .
\end{eqnarray}
Substituting these in~\eqref{SC eps bound} and~\eqref{T'}, we obtain
\begin{eqnarray}
  \nonumber
  T\s' &\geq &
  \frac {\Delta} 
  {\beta (E_1 +\omega)
  -S(E_1+\omega) 
  +V\beta |f_\mathsf{can}|
  +\frac 1 {T\s} J
  +\ln \frac {3 Z\s} 
  {g}}
  \\ \nonumber &\geq &
   \frac {T \Delta} 
  {E_1+\omega 
  -TS(E_1+\omega) 
  +V |f_\mathsf{can}|
  +\frac T {T\s} J
  +T\ln \frac {3 d} 
  {g}}
  \\ \label{T's proves}
\end{eqnarray}
where we have used $Z\s \leq d$ and~\eqref{fed can}.

\subsection{Non-decreasing $S(E)$ case}

In this section we consider the case where the density of states $\Omega(E)$ is non-decreasing for all the energy range of the bath. This is equivalent to not having negative temperature, or $S'(E) \geq 0$.
This usually happens in many-body systems with an infinite Hilbert space dimension, like coupled oscillators, electro-magnetic radiation, or any quantum field theory containing bosons. Clearly, most thermal baths are of this sort.

As far as we know, a many-body system has non-decreasing $S(E)$ if an only if its constituents have infinite-dimensional Hilbert space (e.g. bosons). 
Hence, the content of this subsection does not hold if the constituents of the bath are only spins or fermions.
However, for this last case, one can simplify Result~\ref{R S'>0} by noting that the average energy corresponding to the infinite temperature state constitutes an upper bound for $E_0$, irrespectively of $\xi$.

In the case $S'(E) >0$ we have 
\begin{equation}
  \label {thing we need}
  S(E_1+\omega) \geq S(E_1)
  \ . 
\end{equation}  
Using this and the value of $\omega$ set in~\eqref{omega Ccan} in~\eqref{SC eps bound} gives
\begin{eqnarray}
  \epsilon &\geq &
  \frac 1 {Z\b} 
  \E^{-\beta E_1 +S(E_1) 
  -\sqrt{C_\mathsf{can}}} 
  \, \lambda_{\mathsf{min}}
  \, \frac {d} 3 \ .
\end{eqnarray}
Using the expressions for the canonical~\eqref{fed can} and micro-canonical~\eqref{fed mic} densities of free energy we can write the above as
\begin{eqnarray}
  \epsilon &\geq &
  \E^{-\beta V \left[f_\mathsf{mic} (\frac 1 \xi \ln \frac {2\, d} {3\, g}) - f_\mathsf{can}(\beta) \right]
  -\sqrt{C_\mathsf{can}}} 
  \, \lambda_{\mathsf{min}}
  \, \frac {d} 3 \ ,
\end{eqnarray}
where we have used that the micro-canonical inverse temperature corresponding to $E_1$ satisfies~\eqref{ultime eq E_thr}.

All this procedure can also be applied to~\eqref{T's proves} instead of~\eqref{SC eps bound}, giving 
\begin{eqnarray}
  \nonumber
  T\s' &\geq &
  \frac {T \Delta} 
  {V\! \left[ 
  f_\mathsf{mic} (\frac 1 \xi \ln \frac {2\, d} {3\, g}) 
  - f_\mathsf{can}(\beta) \right]
  +\sqrt{C_\mathsf{can}}
  +\frac {T\, J} {T\s}
  +T\ln \frac {3 d} 
  {g}}
  \\ 
\end{eqnarray}
It is usually the case that $C_\mathsf{can}$ grows with the volume slower than $V^2$. Hence, keeping only leading terms we obtain
\begin{eqnarray}
  \nonumber
  T\s' &\geq &
  \frac {T \Delta} 
  {V\! \left[ 
  f_\mathsf{mic} (\frac 1 \xi \ln \frac {2\, d} {3\, g}) 
  - f_\mathsf{can}(\beta) \right]
  +\frac T {T\s} J
  +T\ln \frac {3 d} 
  {g}}
  \\ \label{T's proves 2}
\end{eqnarray}

\subsection{Extensive bath (Result~\ref{RLT6})}

The above formula can also be achieved by assuming 
\begin{equation}
  \label{scaling of S}
  S(2E,2V) = 2S(E,V)
\end{equation}  
instead of $S'(E) \geq 0$. 
The scaling law~\eqref{scaling of S} implies
\begin{equation}
  S(E,V) = V s(E/V)
  \ ,
\end{equation}
where $s(E/V)$ is the entropy density. Hence, the solution $E_1$ of~\eqref{ultime eq E_thr} satisfies
\begin{equation}
  s'(E_1/V) = 
  \ln\frac {2\, d} {3\, g}
  \ ,
\end{equation}
which implies $E_1 \propto V$.
Relation~\eqref{scaling of S} also implies $C_\mathsf{can} \propto V$ and $\omega \propto \sqrt V \ll V$. Hence, inequality~\eqref{thing we need} is satisfied because $E_1 \propto V$ and
\begin{eqnarray}
  \nonumber
  S(E_1 +\omega) 
  &=&
  V s\!\left( E_1/V + \mathcal O(V^{-1/2}) \right)
  \\ \nonumber &\approx &
  V s\!\left( E_1/V \right)
  \\ &=& S(E_1)
  \ .
\end{eqnarray}
Therefore, with assumption~\eqref{scaling of S} instead of $S'(E) \geq 0$, we can also arrive at~\eqref{T's proves 2}.

\subsection{Radiation-type bath (Result~\ref{RLT7})}
\label{sec:RLT7}

In this section we consider a family of bath systems in which the micro-canonical entropy has the form
\begin{equation}
\label{partI}
  S(E) = \alpha\, V
  \left( \frac E V \right)^\nu
  \ ,
\end{equation}
where $\alpha >0$ and $\nu \in [1/2,1)$ are two constants.
A physical system obeying this is electro-magnetic radiation in the large volume limit. In a box of volume $V$ and $D$ spatial dimensions the value of $\nu$ for the electro-magnetic radiation is $\nu=\frac D {D+1}$. This is thought to be the most entropic bath for its energy and volume\cite{unruh1982acceleration}, and one therefore expects that it is the most advantageous for cooling.

This entropy function has micro-canonical inverse temperature
\begin{eqnarray}
  S'(E) = 
  \alpha\, \nu\, 
  \left( \frac E V \right)^{\nu-1}
  \ ,
\end{eqnarray}
which is always positive (the premise of Result~\ref{RLT6}).
The corresponding heat capacity is
\begin{eqnarray}
  C_\mathsf{mic} (E) = 
  \frac {\alpha\, \nu} {1-\nu}\,
  V
  \left( \frac E V \right)^\nu
\end{eqnarray}
where we have used formula~\eqref{S'' Cmic}. Note that this heat capacity is positive, finite and proportional to the volume.

The solution of equation
\begin{equation}
  S'(E_0) = 
  \frac 1 \xi \ln \frac {2\, d} {3\, g}
\end{equation}
is
\begin{equation}
  E_0 = 
  V \left(
  \frac 1 {\alpha\, \nu\, \xi} 
  \ln \frac {2\, d} {3\, g}
  \right)^{\frac 1 {\nu-1}}
  \ ,
\end{equation}
and gives
\begin{equation}
  S(E_0) = \alpha V 
  \left(
  \frac 1 {\alpha\, \nu\, \xi} 
  \ln \frac {2\, d} {3\, g}
  \right)^{\frac \nu {\nu-1}}
  \ .
\end{equation}
Substituting in the formula for the micro-canonic free energy density~\eqref{fed mic}, and keeping only leading terms in $\xi$, we obtain
\begin{eqnarray}
  \nonumber
  f_\mathsf{mic} 
  (\mbox{$\frac 1 \xi \ln \frac {2\, d} {3\, g}$}) &=&
  \left( 
  \frac {\alpha\, \nu\, \xi} 
  {\ln \frac {2\, d} {3\, g}}
  \right)^{\frac 1 {1-\nu}}
  -
  \frac \alpha \beta
  \left(
  \frac {\alpha\, \nu\, \xi} 
  {\ln \frac {2\, d} {3\, g}}
  \right)^{\frac \nu {1-\nu}}
  \\ &\approx &
  \left( 
  \frac {\alpha\, \nu\, \xi} 
  {\ln \frac {2\, d} {3\, g}}
  \right)^{\frac 1 {1-\nu}}
  \ .
\end{eqnarray}
The canonical free energy density~\eqref{fed can} for the case~\eqref{partI} is
\begin{equation}
  f_\mathsf{can}  (\beta)
  =
  -\frac 1 {\beta\, V}
  \ln \int\!\! dx\ 
  \e^{\beta V \left[
  \frac \alpha \beta x^\nu -x
  \right]}
  \ ,
\end{equation}
where $x=E/V$.
In the large $V$ limit we can use the saddle point method, obtaining
\begin{equation}
  f_\mathsf{can}  (\beta)
  = 
  \left(
  \frac {\alpha} \beta
  \right)^{\frac 1 {1-\nu}}
  \left[
  \nu^{\frac 1 {1-\nu}} 
  -\nu^{\frac \nu {1-\nu}}
  \right]
  \ ,
\end{equation}
which is just a constant.
Therefore, substituting in~\eqref{T's proves 2} and considering the large $\xi$ limit, we obtain
\begin{eqnarray}
  \nonumber
  T\s' &\geq &
  \frac {T \Delta} 
  {V\left[ 
  \frac {\alpha\, \nu} 
  {\ln \frac {2\, d} {3\, g}}\, \xi
  \right]^{\frac 1 {1-\nu}}
  \hspace{-2mm}
  +\frac T {T\s} J
  +T\ln \frac {3 d} 
  {g}}
  \\ \label{special S} &\geq &
  \frac {T \Delta} 
  {V\left[ 
  \frac {\alpha} 
  {\ln \frac {2\, d} {3\, g}}\, \xi
  \right]^{\frac 1 {1-\nu}}
  \hspace{-2mm}
  +\frac T {T\s} J
  +T\ln \frac {3 d} 
  {g}}
  \ ,
\end{eqnarray}
where in the last inequality we have used $\nu \leq 1$.

\subsection{Explicit dependence on time (Result~\ref{RLT8})}

Now, let us assume that the bath is a $D$-dimensional box of volume $V$ containing radiation at temperature $T$. In addition we assume 
\begin{eqnarray}
  t &\geq & 
  \mbox{$\frac 1 v$}\, V^{1/D}\ ,
\\
  t &\geq & 
  \mbox{$\frac 1 u$}\,  \w\ ,
\end{eqnarray}
and substitute in~\eqref{special S}, obtaining
\begin{eqnarray}
  \nonumber
  T\s' &\geq &
  \frac {T \Delta} 
  {(vt)^D \left[ 
  \frac {\alpha} 
  {\ln \frac {2\, d} {3\, g}} 
  \left( J + \frac {T\, J} {T\s} +ut \right)
  \right]^{D+1}
  \hspace{-2mm}
  +\frac {T\, J} {T\s}
  +T\ln \frac {3 d} 
  {g}}
  \\ &\approx &
  \frac {T \Delta} {v^D} 
  \left[ \frac 
  {\ln \frac {2\, d} {3\, g}} 
  {\alpha\, u} 
  \right]^{D+1}
  \frac 1 {t^{2D+1}}
  \ ,
\end{eqnarray}
where in the last line we only kept the leading terms in $t$.

\end{document}